\documentclass{article}
\usepackage[utf8]{inputenc}
\usepackage{graphicx,xcolor,setspace,geometry}
\usepackage{amsmath}
\usepackage{natbib,authblk,bbm,bm}
\usepackage{multirow}
\usepackage{booktabs}

\title{Selecting the Number of States in Hidden Markov Models --- Pitfalls, Practical Challenges and Pragmatic Solutions}
\author{
Jennifer Pohle$^{1}$\footnote{Corresponding author, \texttt{jennifer.pohle@uni-bielefeld.de}}, 
Roland Langrock$^{1}$, 
Floris M.\ van Beest$^{2*}$,
Niels Martin Schmidt$^{2*}$\\
$^1$Bielefeld University, Germany\\
$^2$Aarhus University, Denmark
}
\date{}

\begin{document}
\begin{spacing}{1.2}
\maketitle
\vspace{-2em}








\begin{abstract}
We discuss the notorious problem of order selection in hidden Markov models, i.e.\ of selecting an adequate number of states, highlighting typical pitfalls and practical challenges arising when analyzing real data. Extensive simulations are used to demonstrate the reasons that render order selection particularly challenging in practice despite the conceptual simplicity of the task. In particular, we demonstrate why well-established formal procedures for model selection, such as those based on standard information criteria, tend to favor models with numbers of states that are undesirably large in situations where states shall be meaningful entities. We also offer a pragmatic step-by-step approach together with comprehensive advice for how practitioners can implement order selection. Our proposed strategy is illustrated with a real-data case study on muskox movement.
\end{abstract}
\vspace{0.5em}

\section{INTRODUCTION}

Hidden Markov models (HMMs) are flexible time series models for sequences of observations that are driven by underlying, serially correlated states. Originating from speech recognition, they have found successful applications in various areas such as robotics, finance, economics and social science (cf.\ Chapter 14 of \citealp{zuc16}). Over the last couple of years, they have also emerged as an increasingly popular statistical tool for the analysis of ecological time series data, where they have proven to be natural statistical models for animal movement data \citep{pat16}, general animal behavior data \citep{der16}, capture-recapture data \citep{pra05}, and distance sampling subject to availability bias \citep{bor13}, to name but a few.

In this paper, we discuss order selection in (finite-state) HMMs, i.e.\ how to select the number of states of an HMM, focusing on inference based on maximum likelihood (for alternative Bayesian approaches, see \citealp{rob00}, and \citealp{sco02}). While conceptually order selection appears to be a simple model selection task, in practice it remains a notoriously difficult challenge faced by practitioners. Whether or not order selection involves difficulties very much depends on the purpose of an HMM-based analysis. We distinguish three main types of applications of HMMs: forecasting, classification (in a supervised learning context), and general inference on the data-generating process (unsupervised learning). Order selection is most challenging in the latter case, and we therefore focus on this application specifically in this paper.

In principle, when a maximum likelihood approach is taken it is conceptually straightforward to use information criteria, e.g.\ the Akaike Information Criterion (AIC), the Bayesian Information Criterion (BIC), or variations thereof, to select between models with different numbers of states (cf.\ \citealp{cel08,bac14}, \citealp{zuc16}). However, especially in the ecological literature it has been claimed --- and sometimes demonstrated --- that traditional model selection criteria, and especially the AIC, often lead to the selection of much larger numbers of states than expected {\it a priori} \citep{dea12,van15,lan15,der16,li17}. 

The preference to include many states, particularly in ecological settings, can to some extent be explained by the complexity of such data sets. In addition to the features that actually motivate the use of state-switching models, such as multimodality and autocorrelation (which together are often indicative of the presence of correlated latent states), real data often exhibit further structure, such as outliers, seasonality or heterogeneity between individuals. When neglecting these features in the formulation of an HMM, then additional states may be able to capture this ignored data structure and therefore provide a better model fit than models with a lower, but (biologically) more realistic number of hidden states \citep{lan15,der16,li17}. For analyses where the interest lies on the interpretation of the states, or on the general dynamics of the state process, this behavior of traditional model selection criteria (e.g.\ AIC, BIC) is highly undesirable. As an example, consider the case of animal movement data. In this context, the states can intuitively be interpreted as proxies for the behavioral states of an animal (e.g.\ resting, foraging or traveling), and primary interest usually focuses on identifying the internal and external drivers of behavioral processes. Therefore, in the ideal case, an HMM applied to an animal's movement data can yield a deeper understanding of the behavior of said animal. However, as outlined above, traditional model selection criteria often point to models with large numbers of states which, crucially, may not be biologically interpretable anymore. 

It is our view that, to date, there is no widely accepted procedure that could resolve the practical problems described above related to order selection in HMMs. In this paper, we claim that it is probably not possible to design a one-size-fits-all formal selection method that would reliably yield sensible numbers of states in practice. Therefore, in this paper, we suggest a more pragmatic approach to choose the number of states in practical applications, which takes into account formal criteria for guidance, but also stresses the importance of the study aim, expert knowledge and model checking procedures.

The paper is organized as follows. In Section 2, we lay out the HMM basics, including a brief review of the types of ways in which HMMs are applied, and provide an overview of theoretical and practical aspects related to order selection. In Section 3, we use simulation studies to demonstrate how additional states in an HMM can capture neglected structure in the data, leading standard information criteria to often overestimate the true number of hidden states. In Section 4, we discuss how to pragmatically choose an adequate number of states, and provide practical advice and guidance. Section 5 gives a real-data case study with muskox movement data, illustrating how to implement our pragmatic approach to order selection in HMMs.

\section{HIDDEN MARKOV MODEL BASICS} 

\subsection{BASIC FORMULATION OF HMMS}\label{basic}

\begin{figure}[!htb]
\vspace{1em}
\begin{center}
\includegraphics*[width=0.75\textwidth]{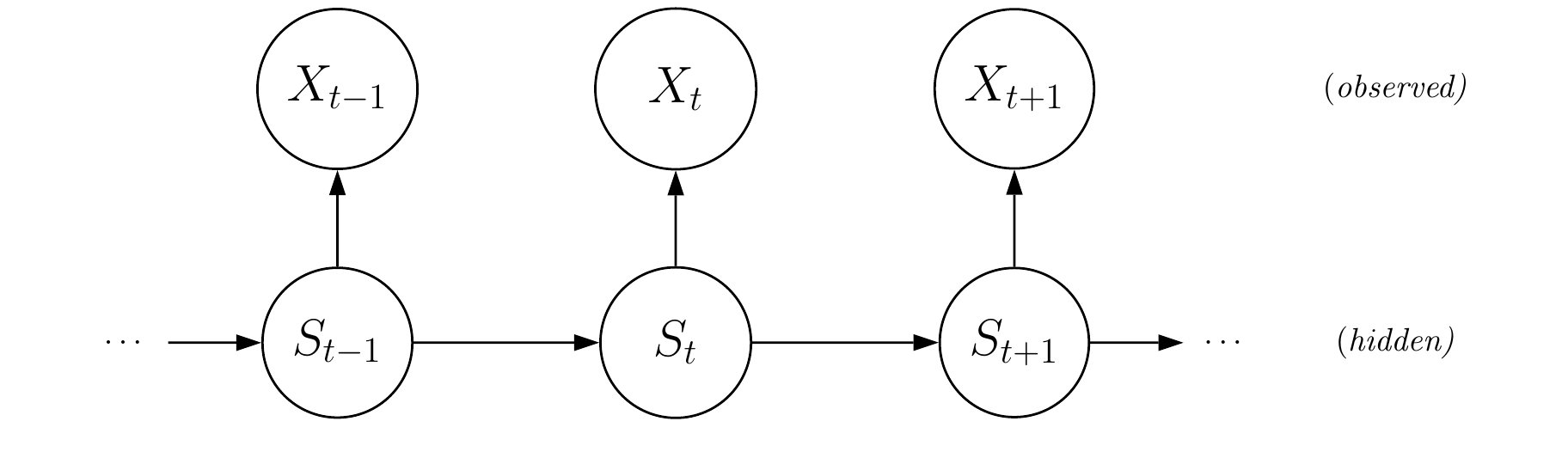}
\caption{Dependence structure of an HMM in its most basic form.}\label{dag}
\end{center}
\end{figure}

An HMM is a doubly stochastic process in discrete time, the structure of which, in its most basic form, is displayed in Figure \ref{dag}. The model assumes the observations, $\{ X_{t}  \,|\, t=1,2,\ldots,T \}$, to be driven by an underlying (unobserved) state sequence, $\{S_{t} \,|\, t=1,2,\ldots,T\}$. In movement ecology, the states are usually interpreted as proxies of the behavioral states of the animal observed \citep{pat16}. The state process is usually assumed to be an $N$-state Markov chain, such that 
$$\Pr(S_{t}|S_{t-1},S_{t-2},\dots,S_{1})=\Pr(S_{t}|S_{t-1}).$$ 
Thus, given the present state, future states are independent of past states. Without loss of generality, throughout the paper we additionally assume the Markov chain to be stationary, unless explicitly stated otherwise. The state-switching probabilities are summarized in the transition probability matrix $\Gamma=\bigl( \gamma_{ij} \bigr)$, where $\gamma_{ij}= \Pr (S_{t}=j \,|\, S_{t-1}=i)$, $i,j=1,\ldots,N$.

In addition to the Markov property, it is usually assumed that the observations are conditionally independent of each other, and of past states, given the current state: 
$$p(X_{t}|X_{t-1},X_{t-2},..,X_{1},S_{t}, S_{t-1},...,S_{1})=p(X_{t}|S_{t}).$$
Here $p$ is used as a general symbol to denote either a probability mass function (if $X_t$ is discrete-valued) or a density function (if $X_t$ is continuous-valued). Thus, the distribution of each observed variable $X_t$, $t=1,\ldots,T$, is completely determined by the current state $S_t$. Together with the Markov property, this assumption gives the basic model structure as directed acyclic graph depicted in Figure \ref{dag}.

In practice, especially with animal behavior data, the observed process will often be multivariate, i.e.\ $\mathbf{X}_t=(X_{1t},\ldots,X_{mt})$. In that case, a commonly made additional assumption is that the $m$ variables are also conditionally independent, given the current state. This is the so-called contemporaneous conditional independence assumption: 
$$p (\mathbf{X}_t \,|\,S_{t}) = \prod_{i=1}^m p(X_{it}\,|\,S_{t}).$$ 
In this basic form, the HMM formulation is completed by choosing the number of hidden states, $N$, and the class(es) of state-dependent distributions (also called emission distributions). In practice, the model parameters will then need to be estimated, which is usually accomplished using either numerical maximum likelihood, the expectation-maximization (EM) algorithm or Markov chain Monte Carlo \citep{zuc16}. In the next section, we discuss in which ways HMMs can be used in applied statistics, and any implications of the particular type of application on the issue of choosing the number of states, $N$.

\subsection{TYPES OF HMM-BASED ANALYSES}\label{typesofapps}

As indicated in the introduction, there are three main types of ways in which HMMs can be applied: forecasting, classification and general inference. (One could argue that a possible fourth type of application is as a tool to make likelihood inference tractable, as demonstrated for example in \citealp{ped11}, or in \citealp{lan13} --- but that type of application is not relevant for the present work.)

First, HMMs can be used for forecasting future values of the observed time series, as typically done in econometric time series analysis (see, e.g., \citealp{ham94}). In these instances, HMMs are not usually regarded as good representations of the true data-generating process, and instead are typically used merely as tools to accommodate features of the observed time series that may be difficult to capture otherwise, using standard time series models (e.g.\ ARIMA- or GARCH-type models). The main purpose of this application of HMMs is usually not to learn something about the data-generating process, but instead to accurately predict future observations. A specific example is given by the application of HMMs to financial share returns \citep{bul11}: here it is not actually reasonable to assume that there is a finite number of (market) states, yet the corresponding models can produce good forecast distributions.

Second, in the machine learning literature, HMMs are often used for (state) classification in a supervised learning context (e.g.\ for speech recognition, gesture recognition, activity recognition, etc.). In these settings, an HMM is trained using data where the underlying states are known, and subsequently applied to new, unlabeled data with the aim to recognize the underlying states. In ecology, HMMs are sometimes applied in this way in order to detect animal behavior states (\citealp{bro14}; see also the discussion in \citealp{leo16}). However, these applications are relatively rare in ecology, since training data, where the states are directly observed, need to be available, which is usually difficult to realize in the field. In this type of application of HMMs, the choice of the number of states is not an issue, as the states and their meaning are predefined.

Third, in an unsupervised context, HMMs are used to learn something about the data-generating process, without defining the role of the states {\it a priori}. Especially in movement ecology, this is the standard way in which HMMs are applied, with the aim of inferring novel aspects related to the behavioral process \citep{mor04,pat09,lan12}. While practitioners may have some expectations regarding the number of states also in the unsupervised context, the identification of the true, or at least a suitable number of states in general still remains a primary aim of empirical studies. Thus, the unsupervised learning context is where order selection in HMMs constitutes the biggest challenge, so it is this case that we focus on in the current paper.

\subsection{MODEL SELECTION FOR HMMS --- THEORY AND PRACTICE}

In practical applications of HMMs, users need to at least (i) specify the dependence assumptions made within the model (typically the Markov property and conditional independence of the observations, given the states), (ii) decide on the class of distributions used for the state-dependent process (e.g.\ normal distributions), and (iii) select the number of states, $N$. In addition, it may be necessary to (iv) decide which covariates to include the model. 

It is our experience that in most practical applications of HMMs, model selection focuses on (iii) and, if applicable, (iv), with (i) and (ii) specified with little or no investigation into the corresponding goodness-of-fit of the resulting models (though there are of course exceptions to this). For the model selection involved in both (iii) and (iv), when a maximum likelihood approach is taken, then information criteria such as the AIC or the BIC are typically used. 

When using the AIC, the focus lies on out-of-sample predictive accuracy. Given a model fitted using maximum likelihood, with corresponding estimate $\hat{\boldsymbol{\theta}}$ for the parameter vector ${\boldsymbol{\theta}}$, the AIC is defined as
$$ \text{AIC} = -2 \log \mathcal{L}(\hat{\boldsymbol{\theta}} | \bm{x}) + 2 p,$$
where $\mathcal{L}(\cdot | \bm{x})$ is the likelihood function given the observed time series $\bm{x}=(x_1,\ldots,x_T)$, and $p$ is the number of model parameters (see \citealp{zuc16}, for details on how to evaluate the likelihood of an HMM). The term $\log \mathcal{L}(\hat{\boldsymbol{\theta}} | \bm{x})$ can be regarded as a simple plug-in estimate of the expected log predictive density (using only the available data to forecast the log likelihood of future data). The log predictive density is one of many examples of a proper scoring rule for assessing predictive accuracy \citep{gne07}, and plays a key role in model selection due to its connection to the Kullback-Leibler divergence \citep{bur02}. Crucially, the plug-in estimator is biased due to overfitting: on average, the model fits the given sample better than an average sample. Under regularity conditions, it can be shown that in the limit (i.e.\ asymptotically, as $T\rightarrow \infty$), this bias converges to $p$. For large sample sizes, $\log \mathcal{L}(\hat{\boldsymbol{\theta}} | \bm{x}) - p$ hence is an approximately unbiased estimator of the expected log predictive density. Akaike \citep{aka73} defined the AIC as $\log \mathcal{L}(\hat{\boldsymbol{\theta}} | \bm{x}) - p$ multiplied by $-2$, the minimization of which is equivalent to maximization of $\log \mathcal{L}(\hat{\boldsymbol{\theta}} | \bm{x}) - p$. In terms of interpretation, the AIC corresponds to an attempt to predict future data as accurately as possible. 

The BIC is defined as 
$$ \text{BIC} = -2 \log \mathcal{L}(\hat{\boldsymbol{\theta}} | \bm{x}) + p \log (T),$$ 
and differs from the AIC in its form only through the increased penalty term (for $T\geq 8$). However, it is derived from a Bayesian viewpoint and aims at identifying the model that is most likely to be true, instead of maximizing prediction accuracy as does the AIC. Under regularity conditions and for large samples, minimizing the BIC is approximately equivalent to maximizing the posterior model probability \citep{sch78}.
Although the BIC was shown to provide consistent estimates of the number of components in independent mixture models under mild conditions \citep{ker00}, for HMMs, consistency of the BIC is not fully established \citep{cel08}. 
More comprehensive accounts on the theoretical background of both AIC and BIC, and also their relation to other model selection concepts, are given in \citet{zuc00}, \citet{bur02}, \citet{gel14} and \citet{hoo15}.

Similarly as the BIC, the integrated completed likelihood (ICL) criterion proposed by \citet{bie01} takes into account model evidence, but additionally considers the relevance of partitions of the data into distinct states, as obtained under the model.  The ICL criterion approximates the integrated complete-data likelihood, i.e.\ the joint likelihood of the observed values $\bm{x}=(x_1,\ldots,x_T)$ {\em and} its associated underlying state sequence $\bm{s}=(s_1,\ldots,s_T)$ using a BIC-like approximation. As the true state sequence is unknown, it is replaced by the Viterbi-decoded state sequence $\bm{\hat{s}}$, i.e.\ the most probable state sequence under the model considered. With $\mathcal{L}_c(\cdot | \bm{x},\bm{\hat{s}})$ denoting the (approximate) complete-data likelihood, the ICL criterion is defined as
$$ \text{ICL} = -2 \log \mathcal{L}_c(\hat{\boldsymbol{\theta}} | \bm{x},\bm{\hat{s}}) + p\log(T).$$ 
As in case of the AIC and the BIC, the model is chosen that leads to the smallest value of the criterion. In the context of HMMs, the simulation studies provided by \citet{cel08} indicate that ICL may actually underestimate the number of states of the HMM in certain scenarios. This can be explained by the preference of the ICL criterion for models where the emission distributions do not strongly overlap. Despite its intuitive appeal in HMM-based clustering tasks, the ICL has not yet widely been used by practitioners working with HMMs, such that experience regarding its practical use is limited.  

Cross-validated likelihood using a proper scoring rule, as suggested in \citet{cel08}, constitutes another alternative, data-driven approach to model selection, which focuses mostly on predictive performance. In the simulation studies provided in \citet{cel08}, cross-validation does not seem to outperform BIC in its ability to find an adequate number of states. Additionally, cross-validation methods can become very computationally intensive, which becomes particularly problematic with the increasingly large telemetry data sets collected in movement ecology, where the estimation of a single model can easily take several hours.

From the theoretical perspective, the behavior of any of these criteria, and of ICL in particular, is still poorly understood. The purpose of this paper is not to contribute to the theoretical development of order selection methods in HMMs, but rather to discuss practical challenges faced by practitioners and how to address them. For an overview of the state of the art of the theoretical development of model selection in HMMs, see Chapter 15 in \citet{cap05}. Additional formal criteria, largely similar in spirit to those detailed above, are discussed in \citet{cel08} and \citet{bac14}. 

In practice, to select the number of states, HMMs with different numbers of states are typically fitted to the data until the AIC/BIC/ICL has reached its lowest value. A key assumption of both AIC and BIC is that the true data-generating process can indeed be represented by one of the candidate models considered. If the candidate models are at least good approximations of the true data-generating process, then this procedure can still be expected to work well \citep{zuc00}. However, when working with complex ecological data, neither an HMM nor in fact any other computationally feasible statistical model can be expected to be a complete representation of the true process, which will typically involve various complex patterns (even if the actual state space is low-dimensional). Thus, intuitively it is clear that undesirable behavior of say AIC or BIC may occur in practical applications of HMMs to complex patterns, and indeed this is what has been observed in various empirical applications in the past, in particular regarding order selection (see, e.g., \citealp{dea12}, \citealp{van15}, \citealp{lan15}, \citealp{der16}). The purpose of this paper is to shed some light on why order selection using these standard criteria is doomed to fail in highly complex empirical applications, and to discuss pragmatic solutions to deal with this challenge. 

\subsection{RUNNING EXAMPLE --- ANIMAL MOVEMENT}

To demonstrate the practical challenges involved when selecting the number of states of an HMM, we use animal movement modeling as a running example throughout this article. Animal movement is one of the most natural and intuitive applications of HMMs, and also constitutes a scenario where overestimation of the number of states is particularly prevalent. While we use this particular area of application of HMMs to fix the ideas, the issues and methods discussed in this paper are not restricted to such data.

In HMM applications to animal movement data, the observed process is usually a bivariate time series comprising step lengths and turning angles between subsequent locations, typically though not necessarily collected using GPS technology. The states of the Markov chain underlying a fitted HMM can then be interpreted as providing rough classifications of the behavioral modes of the animal observed (e.g.\ exploratory vs.\ encamped; \citealp{mor04,pat16}). 

\section{SIMULATION STUDIES}\label{sims}

In this section, we present simulation studies to investigate the performance primarily of the AIC and the BIC, but also of the ICL criterion, when it comes to selecting a suitable number of hidden states. Given the asymptotic equivalence of AIC and cross-validated likelihood \citep{sto77}, we did not implement the latter because of the associated substantial computational effort. ICL in contrast offers a conceptually different, classification-based approach to assessing a model's relative suitability, with a stronger focus on separation of classes, such that it seems worthwhile to be investigated despite the fact that it is not regularly applied in practice.

We showcase scenarios where there is additional structure in the data that is not accommodated within basic HMM formulations. Each type of additional structure considered may be found in real data, and especially within animal behavior data, where the assumptions made with the basic HMM formulation typically are overly simplistic. What will be shown is that in such cases, the misspecification of the model formulation will often be compensated by additional states which, to some extent, ``mop up'' the neglected structure. As a consequence, standard model selection criteria (AIC and BIC) point to models with larger number of states than necessary and adequate. The ICL criterion is much less prone to this behavior, but, as will be detailed below, has other shortcomings.

We examine seven different scenarios: (1) outliers, (2) inadequate state-dependent distributions, (3) temporal variation, (4) individual heterogeneity, (5) a semi-Markov state process, (6) a second-order Markov chain as state process, and (7) a violation of the conditional independence assumption. We start by describing the scenarios in detail (Section \ref{scecon}), and subsequently provide an overview of the results and conclusions drawn from these (Sections \ref{simres} and \ref{discusssimres}).

\subsection{SIMULATION SCENARIOS CONSIDERED}\label{scecon}

\begin{figure}[!htb]
	\centering
	\includegraphics[width=0.5\textwidth]{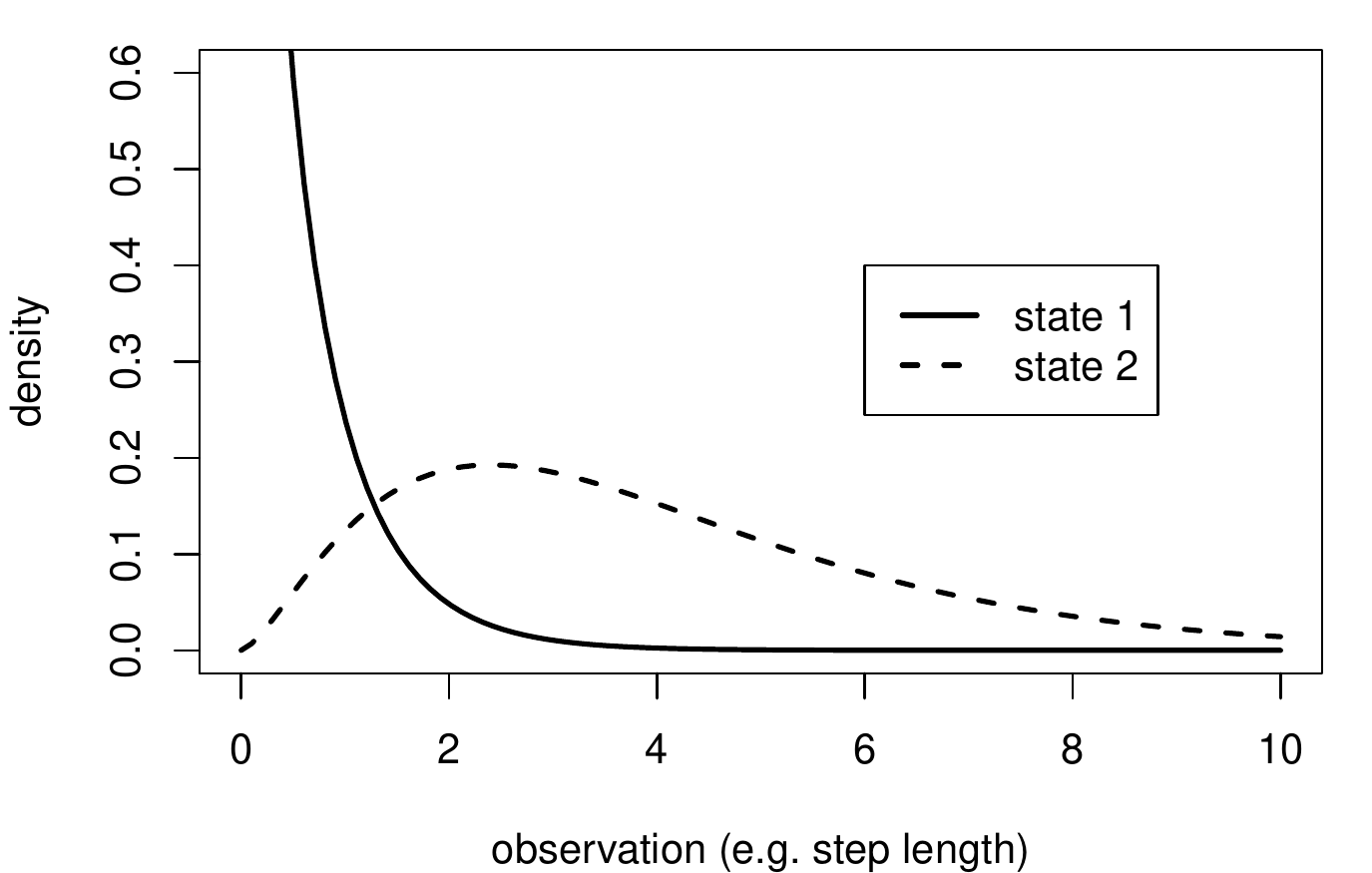}
    \caption{Gamma distributions of the baseline HMM used in the simulation experiments.}\label{fig:basic}
\end{figure}

As a baseline model, we consider a two-state stationary gamma--HMM, i.e.\ an HMM with a univariate, gamma-distributed output variable, with two distinct sets of parameters, one for each state of the underlying stationary Markov chain. This type of model is common for analyzing animal movement data where the observations are step lengths between locations recorded, with the two HMM states corresponding roughly to ``foraging/resting'' and ``traveling'' behavior states, respectively. 
The two state-dependent gamma distributions in the baseline model are shown in Figure \ref{fig:basic}. If not explicitly stated otherwise in the description of the individual scenarios below, then the gamma distribution within the first state has mean$\,=0.5$ and shape$\,=0.7$ (resulting in a strictly monotonically decreasing density function), and the gamma distribution within the second state has mean$\,=4$ and shape$\,=2.5$ (resulting in a density function with mode distinct from zero).

In addition, if not explicitly stated otherwise, then the probability of leaving a state in any given time interval is specified to be 0.1, leading to the transition probability matrix (t.p.m.) 
$$\Gamma=\begin{pmatrix} 0.9 & 0.1\\ 0.1 & 0.9 \end{pmatrix}.$$ 

In each of the seven scenarios considered below, some component of the baseline model formulation will be slightly modified when simulating data. However, crucially, all the scenarios still involve only two genuine states. We will then demonstrate that when not taking the modification into account, i.e.\ when fitting the slightly misspecified basic models to the data, then additional states will be included as per recommendation of model selection criteria, in order to compensate for the inflexibility of the model to otherwise capture the additional structure.

\subsubsection*{Scenario 1 (outliers)}

This first, very simple scenario represents a situation in which some of the data are outliers. In movement ecology, these outliers could for example be due to unusually large  measurement errors (e.g.\ as a result of poor satellite coverage when using GPS tags). The corresponding simulated data are generated using the baseline setup  described above, but subsequently adding uniformly distributed random errors from the interval $[10,20]$ to only $0.5 \%$ of the data points ($25$ data points per sample generated). Intuitively it is clear that these few outlying values may cause the two-state baseline model to have a rather poor goodness of fit, since the two gamma state-dependent distributions may not be able to cover the extreme values without losing accuracy for the non-outlying observations. This can potentially be compensated for by including additional states merely to capture the outlying values.

\subsubsection*{Scenario 2 (inadequate emission distribution)}

While parametric distributions will often provide good approximations of the actual empirical distribution within a state, in practice it will practically never be the case that the true within-state distributions are actual members of some parametric family. In other words, the parametric family being used (e.g.\ gamma or normal distributions) will in general only provide an approximation of the actual data-generating process within a state. In this scenario, we illustrate possible consequences of this for order selection. 

\begin{figure}[!htb]
	\centering
	\includegraphics[width=0.5\textwidth]{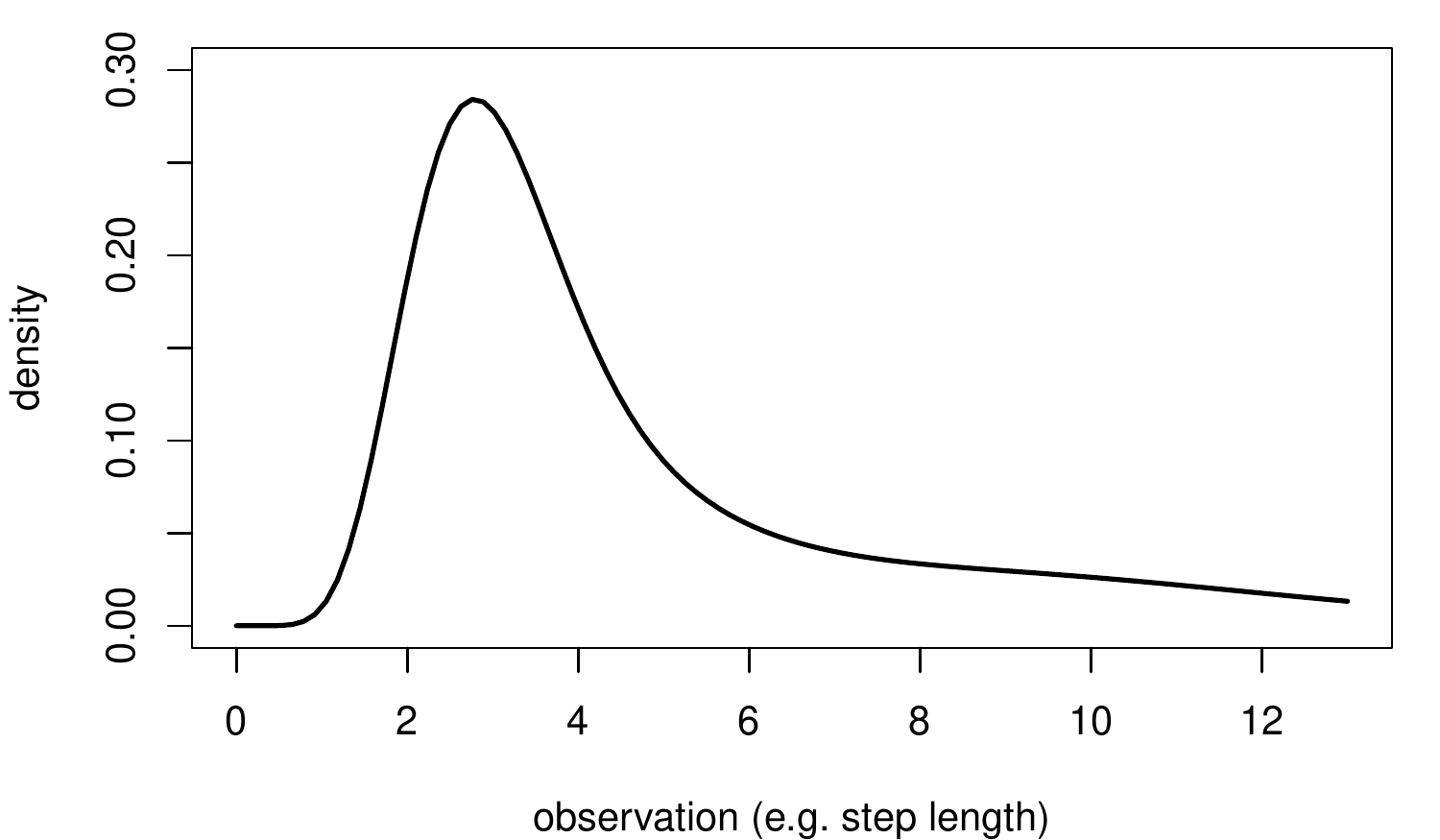}
    \caption{Heavy-tailed emission distribution within state 2, as implemented in Scenario 2.}\label{fig:spline}
\end{figure}

To do so, the observations within the second state were not generated by a gamma distribution with mean 4 and shape $2.5$ anymore, but instead by a similarly shaped but heavy-tailed distribution which we constructed nonparametrically using B-spline basis functions \citep{lan15}. The specific shape of the distribution within state 2 is shown in Figure \ref{fig:spline}.

Clearly, the particular shape of this distribution cannot be fully captured by a single gamma distribution. However, the distribution appears to be such that a two-component mixture of gamma distributions within the second state may be sufficiently flexible to provide a good approximation to the nonparametric distribution. Notably though, a corresponding two-state model, with t.p.m.\ $\Gamma=\bigl( \gamma_{ij} \bigr)$, $i,j=1,2$, and mixture weights $\alpha$ and $1-\alpha$ for the two gamma distributions within state 2, is equivalent to a three-state gamma--HMM with  suitably structured t.p.m.,
$$\Gamma=
\begin{pmatrix}
\gamma_{11} & \alpha (1-\gamma_{11} )& (1-\alpha)(1- \gamma_{11})\\
(1-\gamma_{22}) & \alpha \gamma_{22} & (1-\alpha)\gamma_{22}\\
(1-\gamma_{22}) & \alpha \gamma_{22} & (1-\alpha)\gamma_{22}
\end{pmatrix}.
$$
Therefore, a two-state HMM with a gamma mixture in one of the states can just as well be represented by a three-state simple gamma--HMM. It is thus possible that model selection criteria favor models with more than two states, not because there are more than two genuine (behavioral) states, but because with the additional states it is possible to represent more flexible emission distributions.

\subsubsection*{Scenario 3 (temporal variation)}

In many ecological (and other) time series, there are  clear temporal patterns in the data. We use diel (24 h period) variation as an example but the issue applies to any temporal resolution with variation in the data (e.g.\ seasonal, annual). Diel patterns could be present in the transition probabilities and/or in the parameters of the emission distributions, with the corresponding parameters then being cyclic functions of time.  

\begin{figure}[!htb]
	\centering
	\includegraphics[width=0.5\textwidth]{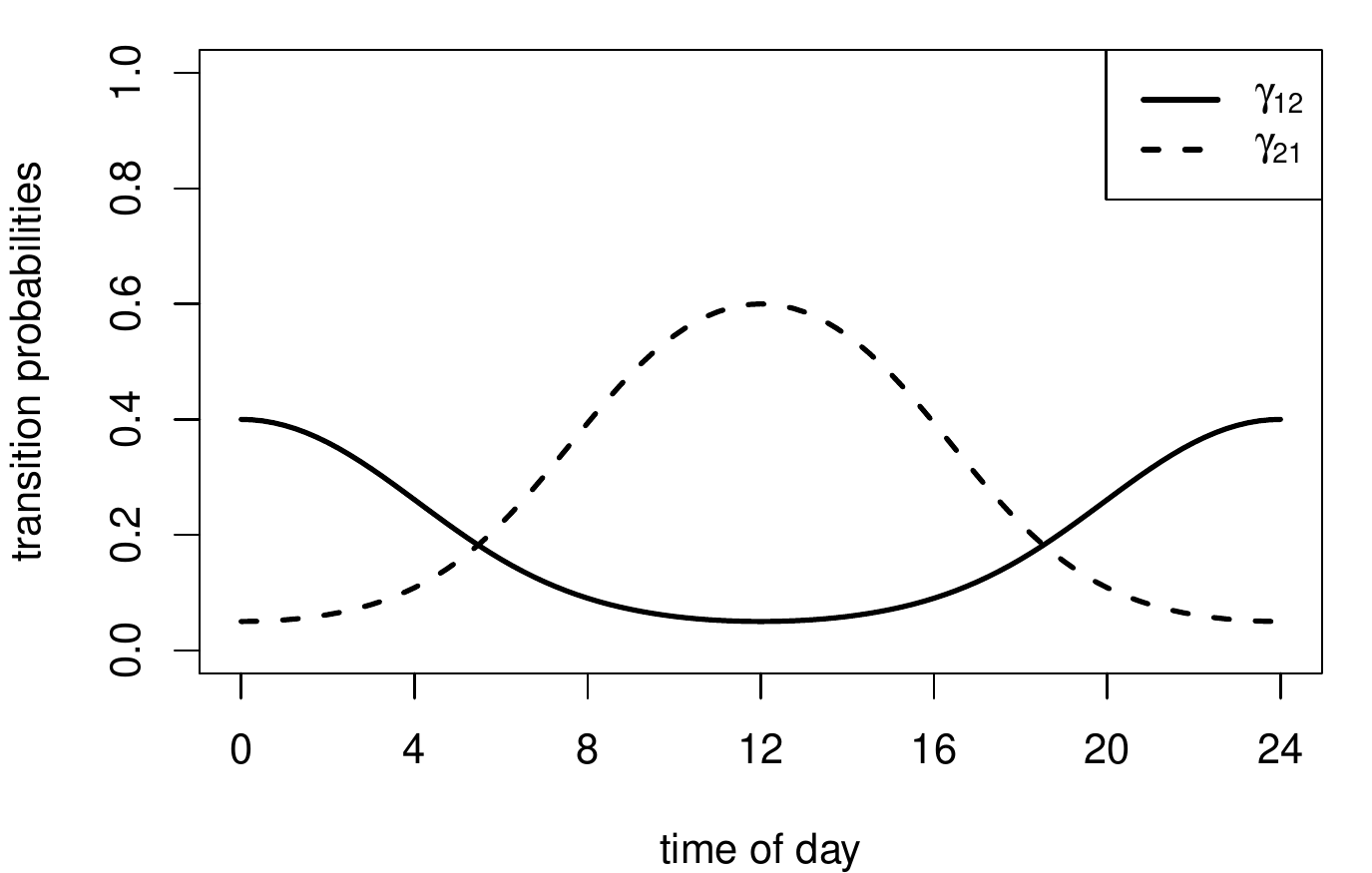}
    \caption{Transition probabilities as a function of time, as implemented in Scenario 3.}\label{fig:season}
\end{figure}

The scenario we consider here is a hypothetical setting with a nocturnal animal that is more likely to be active at nighttime than during the day. The state-switching probabilities will then depend on the time of day. We constructed the transition probabilities using trigonometric functions, with a possible state switch occurring every 15 minutes. The resulting transition probabilities, as a function of time of day, are shown in Figure \ref{fig:season}. 

In the movement ecology literature, analyses including such temporal components were presented for example in \citet{tow16} and in \citet{li17}. Although less clear as for example in Scenarios 2 and 3, it is possible that such a temporal structure in the data, if neglected in the model formulation, may to some extent be captured by additional hidden states.

\subsubsection*{Scenario 4 (individual heterogeneity)}\label{hetero}

When observing more than only one individual, it is natural to assume the individuals to differ, {\it inter alia}, in their personality and fitness, causing heterogeneity between individuals. In the context of animal movement, one could imagine younger individuals to move faster when in an active state, or to occupy certain behavioral states more often, than older individuals. In the former case, the natural modeling approach for such a setting would be to consider individual-specific parameters of the gamma distribution within the active state. For parsimony in terms of the number of model parameters, random effects would typically be used \citep{sch12}. 

Not accounting for such individual heterogeneity within the model formulation, i.e.\ assuming identical within-state gamma distributions across all individuals, could again lead to information criteria favoring models with more than two states: for example, one of the resulting $>2$ nominal HMM states may be capturing the active movement of fitter individuals, while another may be associated with the active movement of less fit individuals, with at least a third state capturing the inactive movement behavior. In our simulations, we simulated $10$ animal tracks of length $500$ each, using a log-normal distribution with the parameters $\mu=\ln 4$ and $\sigma=0.15$ to generate individual means for the (track-specific) gamma emission distribution within the second state.

\subsubsection*{Scenario 5 (semi-Markov state process)}

A first-order Markov chain necessarily implies that the times spent within a state (the so-called dwell times) are geometrically distributed. For example, in the baseline model above, the probability mass function for the time $k$ spent in a state, either state 1 or state 2, is 
\begin{equation}\label{dwd}
p(k)=  0.1 \cdot 0.9^{k-1}, \quad k=1,2,3,\ldots .
\end{equation}
In particular, the mode of the probability mass function of the dwell time is at $k=1$. This implicit consequence of the Markov property will clearly be inappropriate in some applications. To give just one example, in \citet{lan14} it was shown that the distribution of the time beaked whales spend foraging at the bottom of a deep dive is substantially different from a geometric distribution. 

Hidden semi-Markov models are designed to overcome this limitation of HMMs, by explicitly specifying a state dwell-time distribution (e.g.\ a Poisson or a negative binomial), at the cost of a considerable increase in computational effort \citep{gue03}. Notably, any given semi-Markov state process can be arbitrarily accurately represented using a (first-order) Markov state process with expanded state space \citep{lan11}. This implies that when a semi-Markov structure is ignored in the model formulation, then model selection criteria can be expected to point to models with larger number of states than there are genuine (biological)  states, with the model states structured such that the semi-Markov structure is reflected.

The synthetic data in this simulation experiment are generated by the baseline model described above, but replacing the geometric dwell-time distribution within state 2, as given in (\ref{dwd}), by a Poisson distribution with mean $\lambda=3$.

\subsubsection*{Scenario 6 (second-order state process)}

A seemingly similar, yet conceptually different modification of the first-order Markov assumption is to consider higher-order Markov chains for the hidden state process (see, e.g., Chapter 10 in \citealp{zuc16}), thus allowing, at any point in time, the future state to depend not only on the present but also on one or more past states. To the best of our knowledge, this extension of the basic model formulation has in fact not been applied within movement ecology, and indeed may not be that relevant. However, in principle there could be empirical phenomena, not necessarily within ecology, which will indeed exhibit such correlation structures. An example application to eruption times of the Old Faithful geyser is given in \citet{lan12OF}.

Similarly as in case of hidden semi-Markov models, HMMs with underlying higher-order Markov state processes can equivalently be represented as HMMs with first-order Markov state processes with extended state space \citep{zuc16}. If higher-order memory is neglected in the model formulation, then with the identical reasoning as in the previous section, we would expect model selection criteria to favor models with overly complex state architectures.

To demonstrate this issue, the data in this scenario were generated from a second-order Markov chain, determined by the following (time-homogeneous) state-switching probabilities: 
\begin{align*}
P(S_t=2\,|\, S_{t-1}=1,S_{t-2}=1)=P(S_t=1\,|\, S_{t-1}=2,S_{t-2}=2)=0.25;\\ 
P(S_t=2\,|\, S_{t-1}=1,S_{t-2}=2)=P(S_t=1\,|\, S_{t-1}=2,S_{t-2}=1)=0.05.
\end{align*}
This means that switching the state after just having entered it is less likely than when having already stayed in the state for $k>1$ time units.

\subsubsection*{Scenario 7 (violation of conditional independence assumption)}

Conditional independence of the observations, given the states, is one of the key assumptions made in the basic HMM formulation presented in Section \ref{basic}. This assumption is violated if there is additional correlation in the observed time series {\it within a state}. To demonstrate the consequences of not accounting for corresponding structure in the model formulation, in this simulation scenario we consider time-varying mean parameters of the state-dependent gamma distributions, generated using autoregressive processes of order 1, each of them with fairly strong persistence. 

\begin{figure}[!htb]
	\centering
	\includegraphics[width=0.5\textwidth]{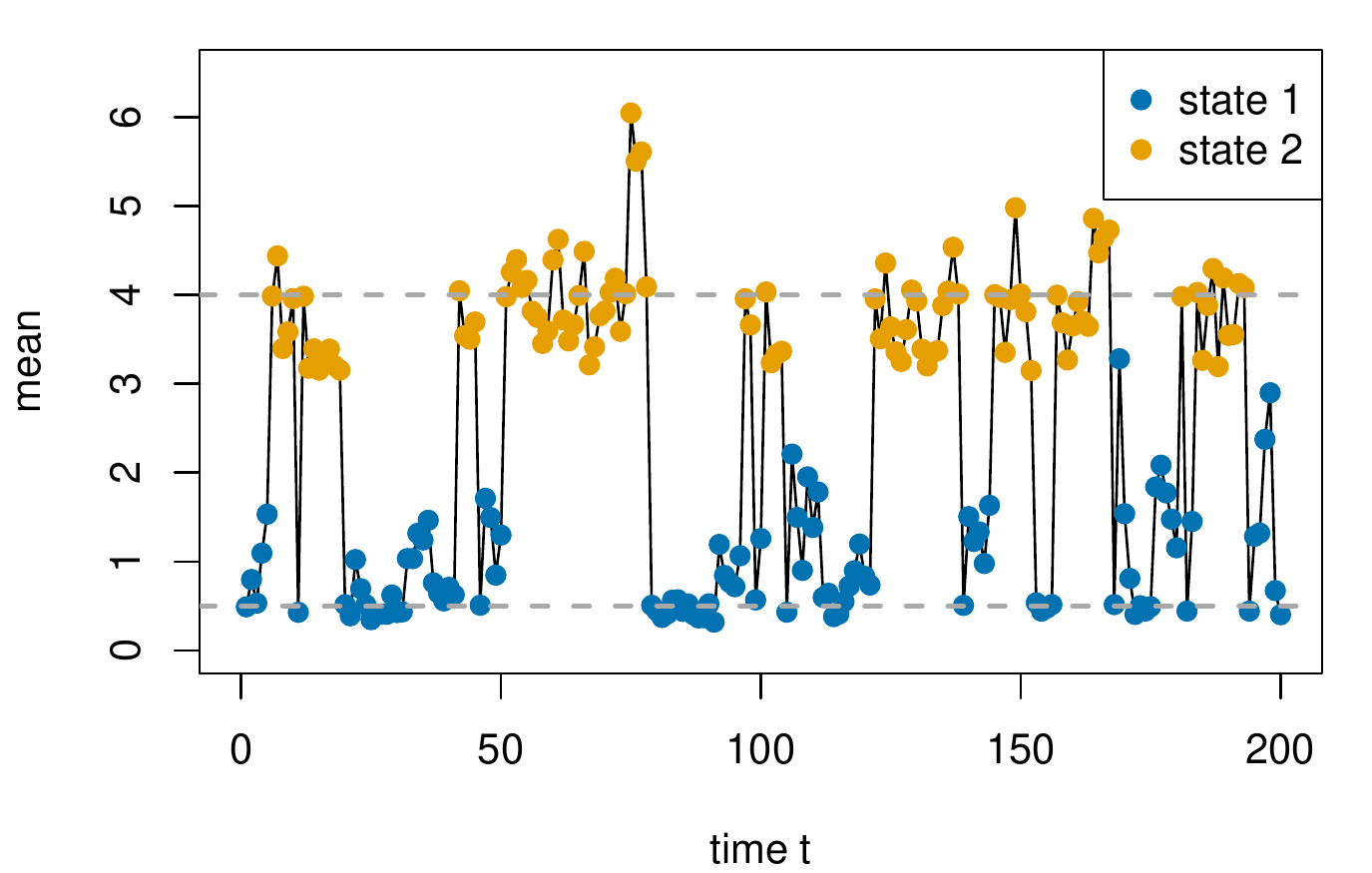}
    \caption{One example (sub-)sequence of gamma mean values generated in Scenario 7.}\label{fig:auto}
\end{figure}

Figure \ref{fig:auto} depicts an example sequence of mean parameters generated in this way, showing only the first 200 time points for clarity. Within state 1, the mean of the gamma distribution fluctuates around $0.5$, while within state 2 it fluctuates around 4. Thus, the state now determines the average level of the mean, but the exact value of the mean depends not only on the state but also on its previous value. For example, in the context of animal movement, the step length of an animal may depend not only on the current behavioral state, but also on the previous speed/step length. Especially at fine temporal resolutions, this would in fact be expected as there will be a certain momentum in the movement. While such fine resolutions are rarely seen with GPS data, they are nowadays commonly seen in analyses of accelerometer data, which can also be tackled using HMMs \citep{leo16}.

With pretty much exactly the same rationale as outlined in Scenario 4 --- just replacing heterogeneity across individuals by temporal heterogeneity within individuals --- it is intuitively clear that, when the additional correlation is not incorporated in the model, model selection criteria may favor models with more states than adequate.

\subsubsection*{Scenario 8 (benchmark, correct model specification)}

As a benchmark, and to demonstrate the performance of AIC, BIC and ICL in idealized settings, we also consider a scenario where the data were generated from exactly the baseline model as described at the beginning of this section.

\subsection{SIMULATION RESULTS}\label{simres}

For each simulation scenario, $100$ data sets were generated, each containing $T=5000$ observations. Within a given scenario, to each of these 100 data sets, simple gamma--HMMs were fitted, i.e.\ models that do not take into account the modification implemented in the given scenario. These slightly misspecified models, with 2--5 states, were fitted to the simulated data sets using numerical maximum likelihood. In each of the individual model fitting exercises, 150 random starting points were considered in the numerical maximization of the likelihood, thus minimizing the risk of finding only local maxima rather than the global maximum of the likelihood function.

\begin{table}[!htb]
\centering
\begin{tabular}{lccccc}		
	simul.\ 		    & 	& \multicolumn{4}{c}{number of hidden states selected} 	\\ \cmidrule{3-6} 
 scenario			    & criterion &	2 (\%)	&	3 (\%)	&	4 (\%)	&	5 (\%)		\\\midrule
1					    & AIC		&	--		&	47		&	49		&	4			\\[0.2em]
(outliers)			    & BIC		&	30		&	70		&	--		&	--			\\[0.2em]
              		    & ICL		&	58		&	42		&	--		&	--			\\ \midrule
2					    & AIC		&	--		&	27		&	60		&	13			\\[0.2em]
(inadequate emission    & BIC		&	--		&	100		&	--		&	--			\\[0.2em]
distribution)           & ICL		&	26		&	71		&	3		&	--			\\ \midrule
3                       & AIC		&	--		&	--		&	57		&	43			\\[0.2em]
(temporal variation)    & BIC		&	14		&	84		&	2		&	--			\\[0.2em]
              		    & ICL		&	100		&	--		&	--		&	--			\\ \midrule
4   				    & AIC		&	3		&	13		&	77		&	7			\\[0.2em]
(individual 		    & BIC		&	79		&	15		&	6		&	--			\\[0.2em]
heterogeneity)          & ICL		&	96		&	1		&	3		&	--			\\ \midrule
5				        & AIC		&	--		&	14		&	74		&	12			\\[0.2em]
(semi-Markov state      & BIC		&	--		&	100		&	--		&	--			\\[0.2em]
process) 			    & ICL		&	100		&	--		&	--		&	--			\\ \midrule
6                       & AIC		&	--		&	1		&	90		&	9			\\[0.2em]
(second-order state     & BIC		&	2		&	98		&	--		&	--			\\[0.2em]
process)                & ICL		&	92		&	2		&	3		&	3			\\ \midrule
7                   	& AIC		&	--		&	--		&	28		&	72			\\[0.2em]
(violation of cond.\	& BIC		&	5		&	95		&	--		&	--			\\[0.2em]
indep.\ assumption) 	& ICL		&	100		&	--		&	--		&	--			\\ \midrule
8					    & AIC		&	37		&	43		&	20		&	--			\\[0.2em]
(benchmark)				& BIC		&	100		&	--		&	--		&	--			\\[0.2em]
              		    & ICL		&	100		&	--		&	--		&	--			\\
		
\bottomrule
\end{tabular}
\caption{Percentages of runs in which the models with 2--5 states are chosen by AIC, BIC and ICL, for all simulation scenarios.}\label{tab:icsim}
\end{table}	

For each scenario, Table \ref{tab:icsim} displays the proportions of the 100 runs in which models with 2, 3, 4 or 5 states were favored by AIC/BIC/ICL, respectively. Both AIC and BIC, which arguably are the most popular tools being used by practitioners, mostly overestimated the number of states that were used to generate the artificial data, in all seven scenarios considered that involve model misspecifications. The AIC performs worse than the BIC, due to the higher penalty on model complexity in the latter. This finding is consistent with the results of \citet{cel08}. The performance of order selection based on AIC and BIC clearly depends, to a large extent, on the severity of the assumption violation being implemented in the simulations. For example, if there were less extreme outliers in Scenario 1, then of course the performance of AIC and BIC with respect to order selection would not be as bad as reported here, and similarly in the other scenarios.

In the simulation experiments considered, the ICL performed very well with regard to selecting the number of states. This can be explained by the tendency of the ICL to favor non-overlapping solutions, i.e.\ HMMs where the state-dependent distributions are clearly distinct. By virtue of the way the state-dependent distributions were defined, this behavior is appropriate in most of the scenarios considered here. The obvious exceptions are Scenarios 1 and 2 --- in the former case due to a third, distinct state-dependent distribution to account for the fairly extreme outliers, and similarly in the latter case, with a third state-dependent distribution accounting for the right tail of the distribution within state 2. While overall positive in the particular simulation setups considered, the focus of ICL on clear separation of the state-dependent distributions is not desirable in general. For example, routine movements of most mammals, that is, movements occurring during an animal’s daily activities, can be dissected into three primary behavioral modes: resting, foraging and traveling. While a traveling mode will typically imply movement patterns clearly distinct from those in the other two modes, it is intuitively clear that the movement metrics (e.g.\ step length) observed in resting and foraging modes, respectively, can actually be very similar, which would typically translate to associated state-dependent distributions that do clearly overlap \citep{leo16}. In corresponding analyses, both within ecology and in other settings, the sensitivity of the ICL with regard to overlapping state-dependent distributions may hinder inference on genuinely distinct modes. These problems have been demonstrated and discussed in \citet{cel08} and in \citet{bac14}. In the Appendix, we provide two additional simulation experiments (Scenarios 9 and 10) showing biologically realistic situations where ICL performs much worse than the BIC due to overlapping emission distributions.

\subsection{DISCUSSION OF THE SIMULATION RESULTS}\label{discusssimres}

It is worth re-iterating that in this paper we focus on analyses where the interest lies specifically in identifying the number of genuine states of the data-generating process, for example because the interpretation of the states matters, or due to a focus on drawing meaningful ecological inference related to the state process. When faced with the choice between a model with the correct number of states, but with a lack of fit in the emission distributions, and a model with too many states that does fit the data well --- cf.\ Scenarios 1 and 2 above, comparing models with two vs.\ such with three states --- we argue that in practice it will then often be preferable to choose the former. Of course, the ideal solution would be to re-formulate the model with the correct number of states such that it does fit the data well (as discussed in more detail below). In any case, it is important to stress that there are of course other situations, particularly those where the aim is to forecast future observations, where it would be preferable to have an essentially incorrect number of states yet a better model fit. Table \ref{biases} illustrates the trade-off between identifying the correct number of states and obtaining unbiased estimators of the emission distributions, by providing summary statistics on the estimates obtained in the 100 simulation runs conducted in Scenarios 1, 5 and 6, respectively. In all three scenarios, the misspecified model with two states led to a slightly biased estimation of some of the parameters of the emission distributions. Including an additional third state --- which here still corresponds to a misspecified model --- led to smaller biases for the emission distributions of the other states within Scenarios 1 and 5, while for Scenario 6, the estimation of the emission distributions was not improved (yet the goodness of fit is also improved; cf.\ Table \ref{tab:icsim}).

\begin{table}[!htb]

	\centering
	\begin{tabular}{ccccccc}
		\multicolumn{7}{c}{} \\			
		\multicolumn{7}{c}{Scenario 1: Outliers} \\ \hline
				&	$\mu_{1}$	&	$\mu_{2}$	&	$\mu_{3}$	&	$\alpha_{1}$	&	$\alpha_{2}$	&	$\alpha_{3}$	\\
true values		&	0.50		&	4.00		&				&	0.70			&	2.50			&					\\[0.5em]
2-state model	&	0.495 		&	4.111 		&				&	0.700			&	2.282			&					\\
				&	(0.016) 	&	(0.060) 	&				&	(0.016)  		&	(0.069)		    &					\\[0.5em]
3-state model	&	0.500		&	3.992		&	15.81		&	0.701   		&	2.523			&	26.33			\\ 
				&	(0.017) 	&	(0.069) 	&	(2.933)	    &	(0.016)		    &	(0.1221)		&	(29.78)		    \\\hline
	\end{tabular}

	\centering
	\begin{tabular}{ccccccc}
		\multicolumn{7}{c}{} \\			
		\multicolumn{7}{c}{Scenario 5: HSMM} \\ \hline
				&	$\mu_{1}$	&	$\mu_{2}$	&	$\mu_{3}$	&	$\alpha_{1}$	&	$\alpha_{2}$	&	$\alpha_{3}$	\\
true values		&	0.50		&	4.00		&				&	0.70			&	2.50			&					\\[0.5em]
2-state model	&	0.506	    &	4.035	    &			    &	0.714		    &	2.620    	    &					\\
				&	(0.025)  	&	(0.052) 	&				&	(0.025) 		&	(0.073) 		&					\\[0.5em]
3-state model	&	0.541		&	3.997		&	0.465		&	0.691			&	2.494			&	0.719			\\
				&	(0.042) 	&	(0.051) 	&	(0.034)	    &	(0.043) 		&	(0.074) 		&	(0.041) 		\\\hline
	\end{tabular}

	\centering
	\begin{tabular}{ccccccc}
		\multicolumn{7}{c}{} \\			
		\multicolumn{7}{c}{Scenario 6: Second-order Markov chain} \\ \hline
				&	$\mu_{1}$	&	$\mu_{2}$	&	$\mu_{3}$	&	$\alpha_{1}$	&	$\alpha_{2}$	&	$\alpha_{3}$	\\
true values		&	0.50		&	4.00		&				&	0.70			&	2.50			&					\\[0.5em]
2-state model	&	0.481		&	 4.015		&				&	0.717			&	2.599			&					\\
				&	(0.023)	    &	(0.060) 	&				&	(0.020)	 	    & 	(0.095)	 	    &					\\[0.5em]
3-state model	&	 0.476		&	4.040		&	2.001		&	0.717			&	2.618			&	0.710			\\ 
				&	(0.025)  	&	(0.088) 	&	(0.608) 	&	(0.022) 		&	(0.103) 		&	(0.414)	    	\\\hline\\
	\end{tabular}
    
\caption{True parameter values and sample mean and standard deviation (in brackets) of the estimates, across the 100 simulation runs, as obtained in simulation scenarios 1, 5 and 6.}\label{biases}
\end{table}

In practice it may be the case that individual assumption violations are much less dramatic in real data than those presented in our simulation setups. On the other hand, with complex empirical data, as typically found in animal behavior time series, we will usually be facing more than just one violation of the assumptions involved in the basic HMM formulation. For example, in the application of HMMs to blue whale dive data reported in \citet{der16}, there were indications of a minor violation of the assumption of contemporaneous conditional independence, strong individual heterogeneity, and a minor lack of fit of the emission distributions. Though these were not explicitly checked, it is likely that also the Markov property and the conditional independence assumption were not fully met. These individual minor deviations of a possible basic HMM being fitted from the true data-generating process may effectively accumulate, such that order selection may be at least as problematic, if not more, than with just a single, yet stronger assumption violation.

Conceptually, it is of course possible to modify the basic HMM, incorporating the additional structure in the model formulation, before tackling the problem of order estimation. For example, in Scenario 2 the reason for the inclusion of additional states, as per recommendation of the model selection criteria considered, is simply the insufficient flexibility of the gamma state-dependent distribution to adequately capture the observations generated within state 2. Such a mismatch between the distributional family employed and the empirical distribution can be detected using residual analyses, as discussed in Section \ref{POS} below. A natural and easy-to-implement remedy would then be to use a finite mixture distribution as emission distribution within state 2. Such a model was implemented for example in \citet{leo16}, in that case effectively merging two states associated with low-activity of eagles. \citet{lan15} discuss the general advantages of using flexible emission distributions, using a nonparametric approach. Considering the misspecifications in some of the other simulation scenarios considered, we note that GPS measurement error can be accounted for within the model formulation \citep{pat08}, mixed HMMs can be used to accommodate individual heterogeneity \citep{alt07}, and semi-Markov or second-order state processes can be implemented to better capture the dependence structure \citep{lan12OF}. When feasible, then improving the model formulation to overcome any substantial lack of fit should be the gold standard. However, each of these modifications is technically challenging and thus difficult to realize for practitioners. In addition, the corresponding models are much more demanding to fit computationally. Especially the estimation of mixed HMMs is well-known to be very computer-intensive \citep{mck15}. When viewed in isolation then each of these extensions will usually still be computationally feasible. However, simultaneously addressing several such patterns will in general be infeasible. Thus, while conceptually it would seem to be most natural to simply overcome the limitations of HMM formulations that cause criteria-based order selection to fail, this is not always a useful strategy in practice. In addition, corresponding highly parameterized models may in fact distract from the actual aim of a study: rather than spending considerable time and effort on technically and computationally challenging non-standard HMM formulations, practitioners will probably seek more pragmatic, goal-oriented ways to overcome the caveats of information criteria in the context of order selection. 

While in some situations the ICL criterion may overcome these problems, we believe that it cannot be seen as a universally applicable solution --- certainly within movement ecology --- due to the potentially undesirable sensitivity to overlapping solutions described above (see also the Appendix). Moreover, the behavior of the ICL criterion is still poorly understood, such that we do not recommend its uncritical use in practice. At the very least, in any given application one needs to carefully consider whether or not clear class separation, as favored by the ICL criterion, is actually expected and/or desirable. 

\section{PRAGMATIC ORDER SELECTION}\label{POS}

Given the difficulties outlined above, we suggest the following pragmatic step-by-step approach to selecting the number of states of an HMM:  
\begin{itemize}
\item[\bf Step 1] decide {\it a priori} on the candidate models, in particular the minimum and the maximum number of states that seem plausible, and fit the corresponding range of models;
\item[\bf Step 2] closely inspect each of the fitted models (starting with the smallest one and subsequently with increasing number of states), in particular by plotting their estimated state-dependent distributions and by considering their Viterbi-decoded state sequences;
\item[\bf Step 3] use model checking methods, in particular residual analyses, but potentially also simulation-based checks, to obtain a more detailed picture of the fitted models, and to validate or invalidate any given candidate model;
\item[\bf Step 4] consider model selection criteria for guidance as to how much of an improvement, if any, is obtained for each increment in the number of states;
\item[\bf Step 5] make a pragmatic choice of the number of states taking into account findings from Steps 2-4, but also the study aim, expert knowledge and computational considerations;
\item[\bf Step 6] in cases where there seems to be no strong reason to prefer one particular model over another (or several other) candidate model(s), results for each of these models should be reported. 
\end{itemize}
The proposed strategy applies only to the unsupervised learning case, where inference related to the state process is of primary interest (cf.\ Section \ref{typesofapps}). The exact procedure we suggest within each step, and the underlying rationale, is detailed below.

\subsection{STEP 1: DECIDING ON CANDIDATE MODELS}

Regarding {\bf Step 1}, it is good practice, not only for HMMs, to restrict model selection only to those candidate models that are plausible, i.e.\ that can be justified {\it a priori}. These are the ``multiple working hypotheses'' conditional on which any subsequent inference is made \citep{bur11}. Considering additional, implausible models increases the likelihood of an undesired selection bias, where, roughly speaking, a model is selected not by merit but because it got lucky, with the data at hand giving a more favorable picture of the model than would be obtained if more data were available \citep{zuc00}. When applying HMMs in movement ecology in particular, we have experienced that it is seldom useful to consider models with more than four states, a) because the biologists, who know their study species extremely well, typically expect 2-4 behavioral states to be present, and b) because models with more than four states typically turn out to be difficult to interpret, and also tend to be numerically much more unstable.

\subsection{STEP 2: INSPECTING THE FITTED MODELS}

The aim of {\bf Step 2} is to develop an understanding of the key patterns picked up in the data by fitted candidate models, and how these relate to biological expectations and the study aim. This often goes a long way in helping to make an informed choice on the number of states. For example, consider the standard HMM formulation that is nowadays routinely applied in movement ecology, where each state is associated with a distinct correlated random walk movement behavior \citep{mor04}. In these settings, the two-state models almost always exhibit the same key pattern, with one state associated with large step lengths and small turning angles (sometimes labeled the ``exploring'' or ``traveling'' state), and the other state associated with much shorter step lengths and many more reversals (the ``encamped'' or ``foraging'' state). It is then usually interesting to see what happens when a third state is included. In many cases, this will lead to either the ``encamped'' state or the ``exploratory'' state splitting up into two states. In the former case, the two states resulting from splitting the ``encamped'' state could for example correspond to ``foraging'' and ``resting'' states, respectively. (We re-iterate at this point that these interpretations are not to be taken too literally, as the HMM states are in general not going to correspond exactly to behavioral states.) When further increasing the number of states, it could for example happen that a state is split but there is no biological reason to distinguish the resulting two states (cf.\ Scenario 2 in Section \ref{sims}), or that the key structure of the model is unchanged, with the additional state explaining only a handful of observations (cf.\ Scenario 1 in Section \ref{sims}). In general, it is our experience that the more states are included, the more difficult it becomes to assign biological meaning to the states. Overall, the purpose of {\bf Step 2} thus is to get an overview of the suitability of the models, in relation to biological expectations and intuition, to the study aim, but also to each other (i.e.\ what additional feature of the data is explained by the model with $N+1$ states that cannot be explained by the model with $N$ states).

\subsection{STEP 3: MODEL CHECKING}

In order to make an informed choice of the number of states, it is important to understand what causes the potential preference for models with many states. This is one of the purposes of {\bf Step 3} (model checking). The main purpose of this step, as in any statistical modeling exercise, is model validation, i.e.\ the assessment if any given candidate model adequately represents the data-generating process. Validation of HMMs via model checking is covered in detail in Chapter 6 in \citet{zuc16}, such that here we focus on the investigation of the role of the number of states.

In the context of HMMs, a model check based on pseudo-residuals \citep{pat09} could for example reveal that a three-state model is preferred over a two-state model by information criteria mainly because the two-state model cannot capture the right tail of the empirical step length distribution in movement data. This would then most likely be due to the inflexibility of the state-dependent step length distribution assumed, rather than a genuine, i.e.\ biologically meaningful, third state responsible for the most extreme step lengths. As a second example, the empirical distribution of the data may be captured accurately by a two-state model, yet the residuals obtained for the two-state model exhibit much stronger autocorrelation than those for the three-state model. This could be an indication of a violation of the dependence assumptions (Markov property and/or conditional independence assumption).  

In case of an identification of weaknesses in the model formulation within {\bf Step 3}, it needs to be decided whether or not any identified assumption violation ought to be addressed as part of the model formulation. For example, if a residual analysis reveals that it is the inflexibility of the state-dependent distribution considered that causes model selection criteria to include additional states, then a more flexible family of distributions can be specified (e.g.\ a mixture), which may then lead to models with simpler state architectures. When considered in isolation –-- i.e.\ when there is only a single such problem to overcome --– then it will often be feasible to formulate and fit corresponding more complex models. However, when dealing with complex time series, for example virtually any ecological time series, there is usually a bit of everything: a minor lack of fit caused by inflexible state-dependent distributions, a correlation structure which is not fully captured by a first-order Markov chain, heterogeneity which is not fully accounted for, etc. Simultaneously addressing all these structural features in the data within an HMM will very quickly lead to heavily parameterized models, the estimation of which might be very unstable, if feasible at all. And even if it is feasible to fit a model that does attempt to take into account all the features present in the data, it will often still be preferable to use a simpler, more stable model which ignores features that {\em are not pertinent to the ultimate aim of the study}. For example, if the focus lies on the effects of environmental covariates on the state-switching dynamics, then a minor lack of fit in the marginal distribution of the observations may not make any difference, and for computational reasons it can then be preferable to stick to the simpler model.

Our rationale is closely related to the concept of tapering effects discussed in \citet{bur02}, where it is stated that biological systems are usually high-dimensional and include a wide range of larger and smaller effects. Very small effects are difficult to uncover within the analysis, although to some extent they do affect the structure of the data set. \citet{bur02} argue that for many studies it will be sufficient to have a low-dimensional representation of the system and therefore to use a simple model which does not include all tapering effects, hence abandoning the idea of identifying a simple ``true model''.

\subsection{STEP 4: CONSIDERING MODEL SELECTION CRITERIA}\label{modsel}

As demonstrated in the simulations in Section \ref{sims}, model selection criteria can be very misleading when it comes to the selection of the number of states of an HMM. Nevertheless, {\bf Step 4} should be implemented, in order to get an overall assessment of any candidate models validated within {\bf Step 3}. If the improvements in say AIC or BIC are very large when increasing the number of states, then this could of course be an indication that the additional states are indeed required. However, as demonstrated in the simulations in Section \ref{sims}, it may just as well be an indication that the additional states merely mop up some structure missing in the model, but have no clear (biological) meaning. At this point it should be clear that it is our view that, when comparing models with different numbers of states, a large difference in AIC, BIC, ICL or any other information criterion alone does not prove that the model with the higher (or lower) number of states is most suitable. However, if the inclusion of additional states substantially improves the AIC or BIC, then this usually indicates problems of the simpler model. What exactly these problems may be then needs to be investigated further.

\subsection{STEP 5: PRAGMATIC ORDER SELECTION}\label{pos}

At this point, a lot of information has been gathered which should facilitate the selection of an appropriate number of states. As detailed in Section \ref{modsel}, it is our view that one cannot in general rely on model selection criteria. Instead, the selection of the number of states should take into account:
\begin{itemize}
\item realism of the fitted candidate models, assessed using expert knowledge (as per {\bf Step 2});
\item the results from model checking, in particular in relation to the study aim ({\bf Step 3}); 
\item model selection criteria for guidance ({\bf Step 4}); 
\item potentially computational considerations, if relevant (using the simpler of two equally reasonable candidate models; cf.\ {\it Occam's razor}).
\end{itemize}
As a consequence, the selection of the number of states will necessarily be somewhat subjective. However, we believe that this is the best that can be done in the unsupervised learning context (cf.\ Section \ref{typesofapps}). Furthermore, we have experienced that a thorough implementation as detailed in {\bf Steps 1-4} will usually make it fairly easy to pick a suitable $N$. First of all, the inspection of the fitted models within {\bf Step 2} will often leave only two or perhaps three candidate models as reasonable contestants (see for example \citealp{der16}, but also our real data case study in Section \ref{casestudy}). Together with thorough model checks as implemented within {\bf Step 3}, the advantages and disadvantages of the remaining contestants often become apparent, such that a final pragmatic choice of $N$ can be made. In movement ecology, some sort of pragmatic order selection has in fact been implemented in several contributions, including for example \citet{mor04}, \citet{dea12}, \citet{van15} and \citet{der16}. 

As in any other statistical modeling exercise, order selection in HMMs can of course be an iterative process. That is, after going through {\bf Steps 1-4}, it may become clear that alternative model formulations need to be considered, e.g.\ modifying the dependence assumptions, such that one needs to return to {\bf Step 1}.

\subsection{STEP 6: REPORTING OF SEVERAL MODELS}

While it will sometimes be straightforward to make a pragmatic choice as described in {\bf Step 5}, there will certainly also be cases where two or more candidate models may seem more or less equally suitable after following the steps detailed above. In those cases, it is our view that best scientific practice is to report the results of all suitable models (as recommended also by \citealp{bur02}; and quoting \citealp{cha12}, ``the active realist ideal is not truth or certainty, but a continual and pluralistic pursuit of knowledge''). In the context of order selection for HMMs, this translates to acknowledging uncertainty and presenting results accordingly. However, at least within movement ecology, this is hardly ever done (but see \citealp{mor04}, \citealp{lan15}, \citealp{der16}). 

\section{CASE STUDY: MUSKOX MOVEMENT}\label{casestudy}

We demonstrate the workflow of our suggested pragmatic approach to order selection using movement data collected for a single adult female muskox in east Greenland, which was observed for a period of nearly 3 years. The raw data set comprises $t=25103$ hourly location observations obtained using GPS collars (including about 1\% missing locations). These were used to calculate the hourly step lengths and turning angles, i.e.\ the metrics that are commonly used when applying HMMs in movement ecology \citep{mor04,pat16}.

In this case study, we focus on inference on the (behavioral) state-switching dynamics within an unsupervised learning context (cf.\ Section \ref{typesofapps}). We considered fairly basic HMMs, using gamma distributions for the step lengths and von Mises distributions for the turning angles. Some step lengths are exactly equal to zero, such that we allowed for an additional point mass on zero. We assumed step lengths and turning angles to be conditionally independent, given the states (contemporaneous conditional independence). 

Following Section \ref{POS}, we first ought to decide {\it a priori} on the minimum and maximum number of states that seem plausible ({\bf Step 1}). The most dominant behavioral states are expected to be ``resting/ruminating'',  ``feeding'' and ``moving''. These are quite typical for most ruminating/large animal species. Thus, from a biological perspective, three or four behavioral states for muskoxen seem most reasonable \citep{sch16}. Since it is not clear if exactly these actual behavioral modes will manifest themselves for the given data set, and in particular at the time scale considered (hourly observations), we included models with 2-5 states in our candidate set. Parameter estimation was carried out using numerical maximum likelihood as implemented in the R package moveHMM \citep{mic16}. For each model considered, $50$ randomly chosen starting values for the parameters were used in order to minimize the chances of missing the global maximum.  

\begin{figure}[!htb]
\vspace{1em}
\begin{center}
\includegraphics*[width=0.95\textwidth]{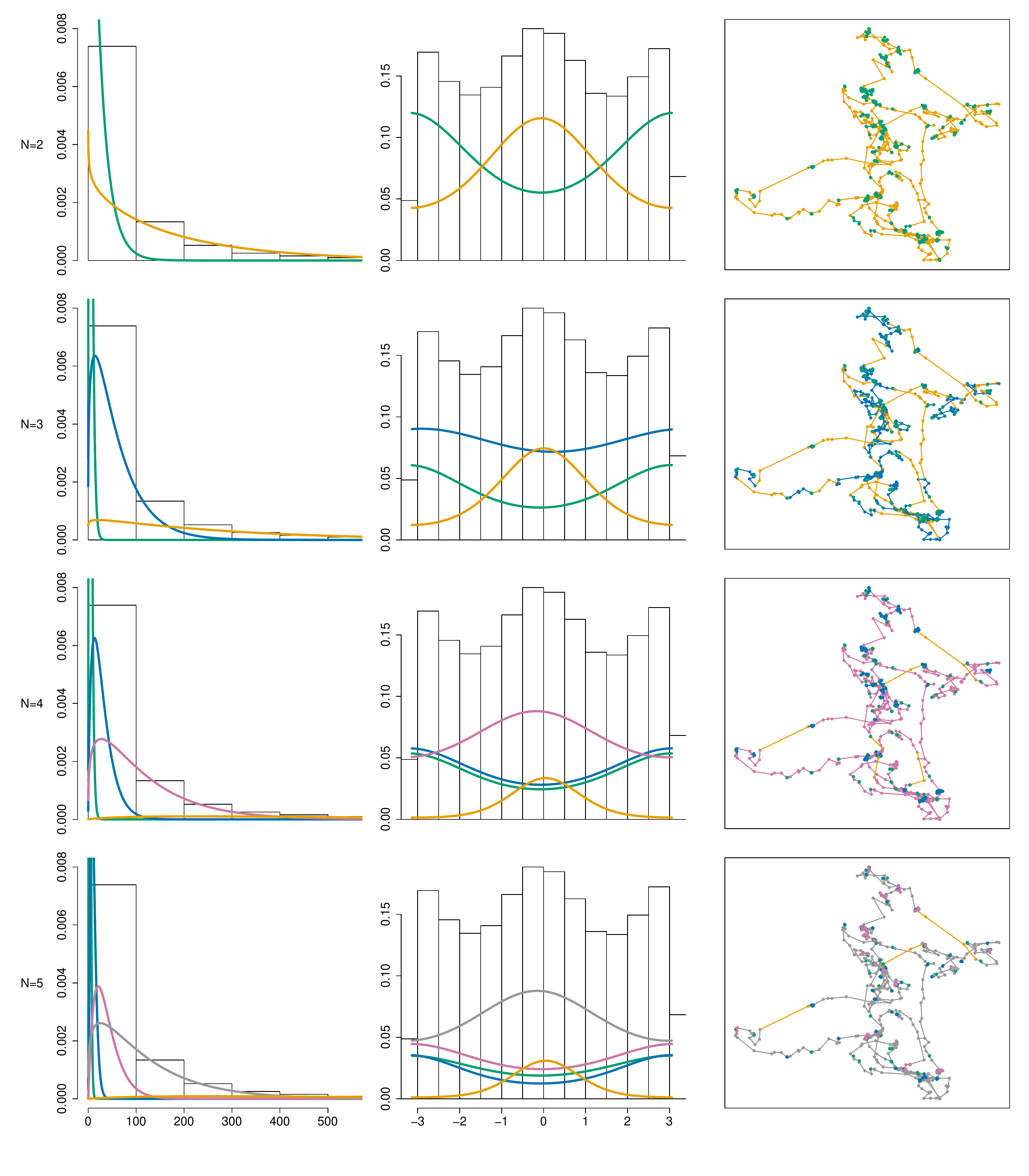}
\caption{Estimated state-dependent distributions for models with 2--5 states (one row for each model, gamma step length distributions in the left column, von Mises turning angle distributions in the middle column), and associated decoded state sequences (right column). }\label{fig:mox}
\end{center}
\end{figure}

For {\bf Step 2} (inspecting the fitted models), Figure \ref{fig:mox} displays, for each model, the state-dependent distributions estimated for the step lengths and the turning angles, respectively, as well as a Viterbi-decoded sequence of states for a subset of the observed time series of locations. In the plot, the state-dependent distributions are weighted with the proportion of time the corresponding state was occupied, as determined by the Viterbi algorithm. For the two-state model, the fitted distributions exhibit the standard pattern typically found in these movement models, with one state involving large steps and directed movement and the other state much smaller steps and many reversals. This is confirmed by the decoded state sequence. 

When adding a third state, the distributions are still fairly easy to interpret. Interestingly, the state associated with the smallest steps now involves hardly any movement (Figure \ref{fig:mox}). This corresponds well to the (biological) expectation of a ``resting''/``ruminating'' state. With the low-activity state focusing on this virtually stationary behavior, the other two states now provide a more nuanced differentiation of active movement behavior. More precisely, one of the more active states involves relatively long steps (mean$\,\approx 286$ meters) on average and few turnings (directed movement), while the other active state involves moderate steps (mean$\,\approx 59$ meters) and many turnings. The former pattern is strongly indicative of a movement mode without any clear foraging activity, while the latter suggests foraging behavior with small localized search movements. These results make biological sense and agree with previous empirical findings \citep{sch16}.

The four-state model can most easily be compared to the two-state model described above. Roughly speaking, when going from two to four states, both the more active and the less active modes are split into two separate states. While the split of the inactive state seems biologically sensible (see above discussion), it is much less clear if the distinction between ``moving'' and ``(long-distance) traveling'' is really necessary and useful. In particular, the latter state is active only about 6\% of the time, according to the Viterbi-decoded state sequence. Although such infrequent long-distance traveling behavior is possible from a biological perspective (e.g.\ fleeing or displacement behavior following a predation or disturbance event), this is unlikely to be the case for muskoxen in this part of east Greenland \citep{sch16}. As discussed in Section \ref{discusssimres}, in such a situation it would be natural to try to effectively merge these states via the use of a more flexible emission distribution, example using a mixture emission distribution.

While the four-state model is still interpretable (at least to some extent), adding a fifth state does not add any clear value and muddles interpretation (Figure \ref{fig:mox}). In the corresponding model, there are now two states which involve hardly any movement (with hourly step length means of 3.4 and 10.8 meters, respectively), which could simply be artefacts of temporal variation in GPS measurement error, which is expected to be ca.\ 10 m for the GPS collar used here.

\begin{figure}[!htb]
\vspace{1em}
\begin{center}
\includegraphics*[width=0.7\textwidth]{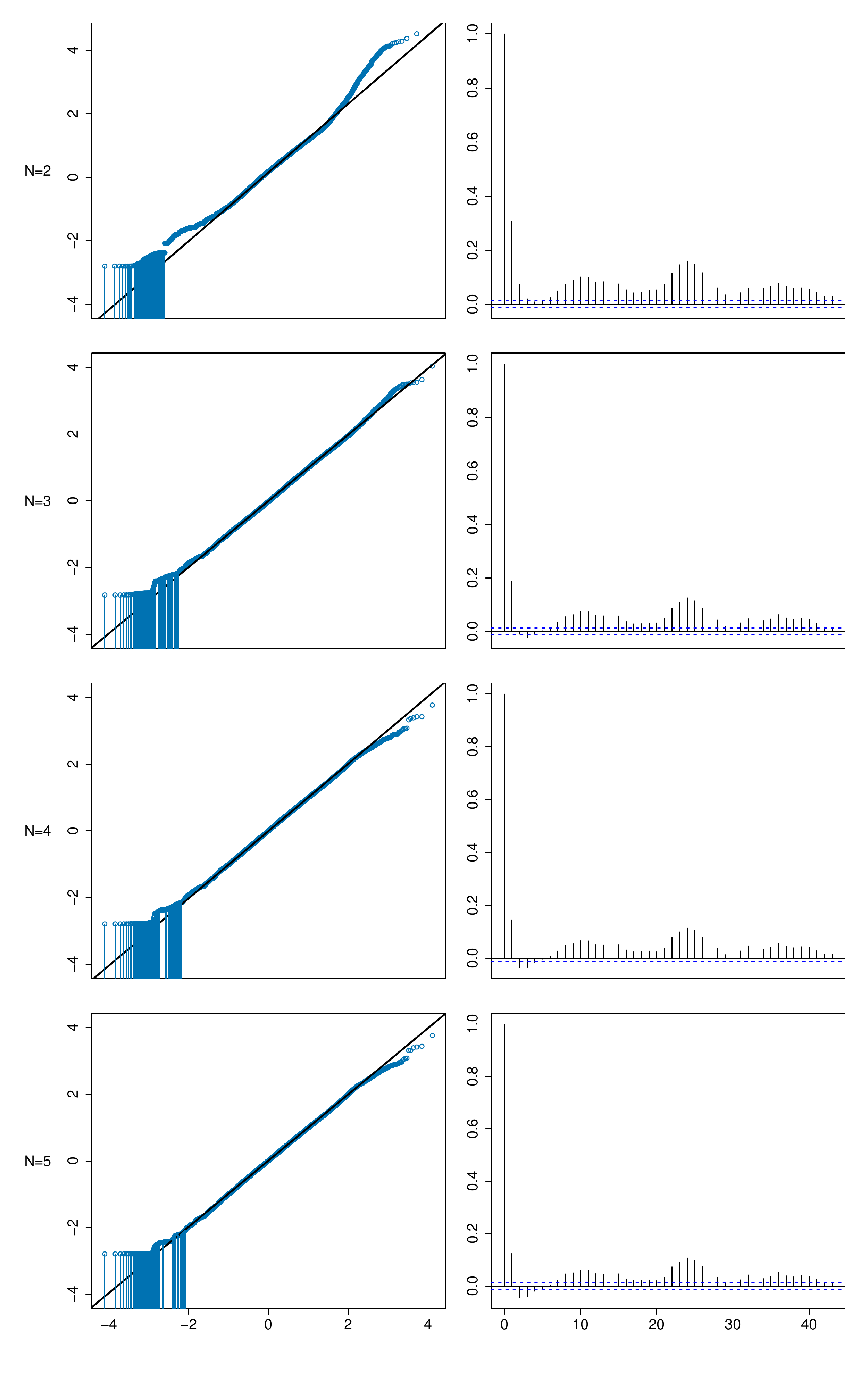}
\caption{Quantile-quantile plots and autocorrelation functions of the pseudo-residuals obtained for the four different models fitted (with 2-5 states, from top to bottom) to the muskox movement data.}\label{fig:check}
\end{center}
\end{figure}

We restrict model checking ({\bf Step 3}) to the pseudo-residuals of the step lengths, since turning angles are not as easily amenable to a residual analysis due to their circular nature \citep{lan12}. Figure \ref{fig:check} shows the quantile-quantile plots of the pseudo-residuals and their autocorrelation functions, for each estimated model. The pseudo-residuals of the two-state model indicate a fairly substantial lack of fit in both the upper and the lower tail, while the models with three states or more appear to provide a reasonable fit of the marginal distribution of the step lengths. However, for all models considered, the respective sample autocorrelation functions (ACFs) clearly do indicate another source of lack of fit, namely that there is some diel variation in the data, which is not taken into account within our models. As we have seen in the simulation experiments (Section \ref{sims}), this may already cause model selection criteria to point to models with more states than necessary and biologically sensible. And indeed, the sample ACF of the residuals obtained in case of the five-state model shows a less strongly marked diel pattern than the ACF of the residuals for the two-state model, despite both models not explicitly taking this feature of the data into account.

For {\bf Step 4} (considering model selection criteria), Table \ref{tab:icox} displays the AIC, BIC and ICL values for each model fitted. Both AIC and BIC favor the five-state model. In fact, both criteria are further improved when considering even more states. We tried up to nine states, and from all models considered the nine-state model was deemed optimal by both AIC and BIC. This could indicate that there is much more structure in the data than assumed by a basic HMM as the ones fitted here. Indeed, muskox move in a highly seasonal and dynamic environment (the Arctic) where environmental conditions can change rapidly over time (e.g.\ weather patterns) and space (e.g.\ heterogeneity in availability of vegetative cover). The movement patterns of muskoxen could therefore be too complex to capture with a simple three-state HMM with strong dependence assumptions, though we cannot draw any firm conclusions at this point. Notably, the ICL does not point to the most complex model being fitted but to the four-state model, which as detailed above seems largely appropriate.

\begin{table}
	\centering
	\begin{tabular}{ccccc}
		\multicolumn{5}{c}{} \\	\hline
			no.\ states	& no.\ parameters	&	AIC			&	BIC				& ICL				\\ \hline
			2		&		12		    &	350199.3		&	350296.7  		& 354829.3 			\\
			3		&	 	21			&	345285.4		&	345455.8  		& 351544.5			\\
			4		&		32			&	343404.9		&	343664.6  		& {\bf  350159.9}	\\
			5		&		45			& {\bf 342782.0}	& {\bf 343147.2} 	& 351247.7			\\ 				
	\end{tabular}
		\caption{AIC, BIC and ICL values obtained for the different models fitted to the muskox movement data. The models selected by the different criteria are highlighted in bold face font.}\label{tab:icox}
\end{table}

At this point, there are basically two options, and it depends on the aim of the study which of the two should be pursued. First, it may be relevant to explicitly account for the diel patterns exhibited by the muskox within the model, say when investigating the state-switching dynamics in relation to internal and external drivers (a corresponding application is described, for example, in \citealp{li17}). In that case, one needs to return to {\bf Step 1} and formulate corresponding candidate models, then proceeding with {\bf Steps 2-6}. It could also be worthwhile to investigate if more flexible emission distributions, e.g.\ mixtures, would substantially improve the fit of the models with only two or three states. Second, it may be the case that the diel variation and any minor lack of fit of the emission distributions can be neglected because it does not interfere with the study aim. 
For example, the primary interest may lie in identifying the spatial regions in which an animal is most likely to forage during a specific time window in which diurnal or environmental variation is low (i.e.\ high Arctic summer with 24 hours of daylight and abundant vegetation). In such a case, whether or not the exact correlation structure of the state process is captured will likely have very little influence on the state decoding, such that it may be preferable to stick to the simpler models.

As this case study is supposed to merely illustrate the workflow of the pragmatic order selection approach suggested, we do not pursue the former of the two options. If the latter route is taken, then taking into account the findings from above, it is clear that either the three-state or the four-state model constitutes a pragmatic yet justified choice ({\bf Step 5 \& 6}).

\section{DISCUSSION}

In this study, we have demonstrated why model selection criteria are problematic with respect to choosing the number of states of an HMM applied to complex real data within an unsupervised learning framework. More specifically, AIC and BIC tend to favor models with too many states, since any structure in the data that is neglected in the model formulation will, to some extent, be mopped up by additional model states that do not have a clear interpretation anymore. Since the performance of model selection criteria strongly depends on the severity of any deviations of the HMM being fitted from the {\it unknown} true process, it is intuitively clear that no one-size-fits-all objective and universally applicable criterion can be developed for order selection in HMMs. The ICL criterion appears to overcome several of the problems associated with the more established AIC and BIC, yet it does not come without its own limitations, namely a sensitivity to overlapping state-dependent distributions. While in some applications a clear separation of latent classes may be appropriate, this is not generally the case, especially in animal movement modeling. Thus, we proposed a pragmatic step-by-step approach to order selection which, while lacking objectivity, we believe is the best possible practical solution.

In this manuscript, we focused exclusively on inference based on maximum likelihood, and did not investigate whether similar problems arise under a Bayesian paradigm. In principle, the possibility to use priors to effectively exclude models with undesirably large numbers of states seems appealing. \citet{rob00} proposed reversible jump Markov chain Monte Carlo algorithms in particular for choosing the number of states. Notably, those authors also find that ``adding structure [...] to a model pushes the posterior distribution of $k$ towards smaller values'' (where $k$ is the number of states), which indicates that the conceptual problem of additional states mopping up neglected structure arises in both inferential frameworks. Overall, we do believe that the Bayesian framework may in principle offer opportunities for formalizing the concept of pragmatic order selection, but also that a certain level of subjectivity is unavoidable --- to use priors to enforce a small number of states in our view effectively just shifts the problem. Further comprehensive overviews of Bayesian inference in HMMs are given in \citet{sco02}, \citet{cap05} and \citet{fru06}.

Many of the key principles brought forward in this paper have previously been presented in the area of cluster analysis, in particular by Hennig (2015). In the context of finding the ``true'' number of clusters for a given data set, Hennig argues that there is a ``misguided desire for uniqueness and context-independent objectivity''. This, as should be clear from the presentation of our suggested pragmatic approach, is very much the same view that we have on the issue of order selection in HMMs. Similarly as in cluster analysis, there may also be different concepts of what constitutes a ``true'' state. Corresponding definitions may be based on the data alone, on external {\it a priori} information, or on HMM fitting results, and in general are to be seen in relation to the study aim (Hennig, 2015). The key point for us, as also pointed out in Hennig (2015), is that the individual researcher's modeling decisions, and in particular the rationale underlying the selection of the number of states, need to be made transparent. While the pragmatic approach to order selection presented in this paper clearly depends on subjective decisions made by the researcher, a corresponding analysis nevertheless will be as scientific, if not more scientific, than any allegedly objective choice of the number of states.

The somewhat problematic notion of a ``true'' state is in fact exacerbated within movement ecology, where the meaning of a state may strongly depend on the time interval the Markov chain operates on, and hence on the sampling protocol applied when collecting location data. Nevertheless, based on our experience there is a strong desire within the movement modeling community to use statistical models to infer actual behavioral states from tracking data. Thus, in this paper, we have focused on corresponding situations where practitioners do seek meaningful states. In those cases, pragmatic order selection as detailed in Section \ref{pos} will often lead to the choice of a model with a small number of interpretable states, at the expense of a lack of fit of the corresponding model, and potentially biased estimation in particular of the state-dependent distributions. The muskox case study illustrates this trade-off: here a model with five or more states, while superior in the goodness of fit, would not be useful in say determining drivers of resting and foraging behavior, simply because there is no clear correspondence between model states and behavioral states. In other situations, e.g.\ when interest lies in forecasting future observations, it can of course be preferable to use a larger model, essentially with too many states, which fits the data well.

Overall, the selection of the number of states clearly is an important yet challenging issue, which requires statistical expertise (when applying model selection and model checking tools) and modeling experience, but also a good understanding and intuition of the data and research question at hand (in order to arrive at a sensible choice of the number of states). Within statistical ecology, this underlines the need for statisticians and ecologists to closely collaborate in all stages of an analysis.

\section*{Appendix --- additional simulation study}

In the main article we state that the ICL is sensitive with regard to overlapping state-dependent distributions. These problems have been demonstrated and discussed for example in \citet{cel08} and in \citet{bac14}. However, the simulation experiments described in the main article were designed as to highlight the practical problems involved specifically when using AIC or BIC for order selection, as these arguably are the most popular tools being used by practitioners. As such, they do not explicitly focus on scenarios with strongly overlapping state-dependent distributions, which explains the good performance of ICL in these experiments. To explicitly demonstrate that the ICL can indeed be problematic in scenarios with overlapping state-dependent distributions --- as stated in the main manuscript --- in this appendix we provide two additional simulation experiments. Since the ICL criterion is calculated using the Viterbi-decoded state sequence $\hat{s}$ instead of the ``true'' but unobserved state sequence, uncertainty in the Viterbi sequence also influences the performance/accuracy of the ICL. 

We consider two additional simulation experiments, both with gamma state-dependent distributions defined by their state-dependent mean and shape parameters, respectively:
\begin{itemize}
\item {\bf Scenario 9}: $T=1000$, mean $=\begin{pmatrix} 0.5 \\ 1.5 \\ 3 \end{pmatrix}$, shape $=\begin{pmatrix} 2 \\ 3 \\ 4 \end{pmatrix}$, 
$\Gamma=\begin{pmatrix} 
0.9 & 0.05 & 0.05 \\
0.05 & 0.9 & 0.05 \\
0.05 & 0.05 & 0.9 \\
\end{pmatrix}$



\item {\bf Scenario 10}: $T=2000$, mean $=\begin{pmatrix} 5.5 \\ 3.0 \\ 1.0 \end{pmatrix}$, shape $=\begin{pmatrix} 12.0 \\ 4.0 \\ 1.5 \end{pmatrix}$, 
$\Gamma=\begin{pmatrix} 
0.8 & 0.1 & 0.1 \\
0.1 & 0.8 & 0.1 \\
0.1 & 0.1 & 0.8 \\
\end{pmatrix}$
\end{itemize}

\begin{figure}[!htb]
\centering\includegraphics[width=0.6\textwidth]{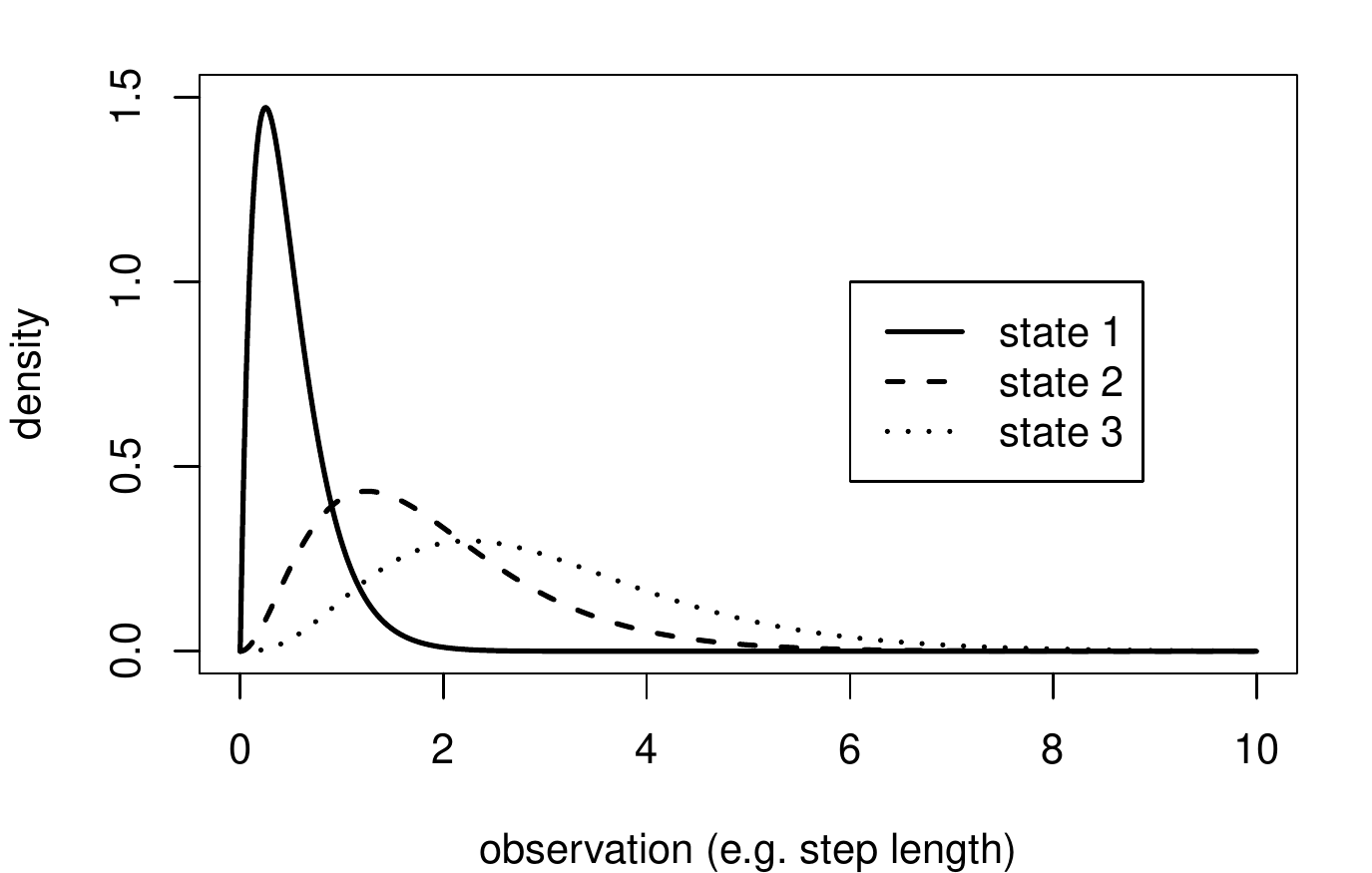}
\caption{Densities of the state-dependent distributions specified in Scenario 9.}\label{fig9}
\end{figure}

\begin{figure}[!htb]
\centering\includegraphics[width=0.6\textwidth]{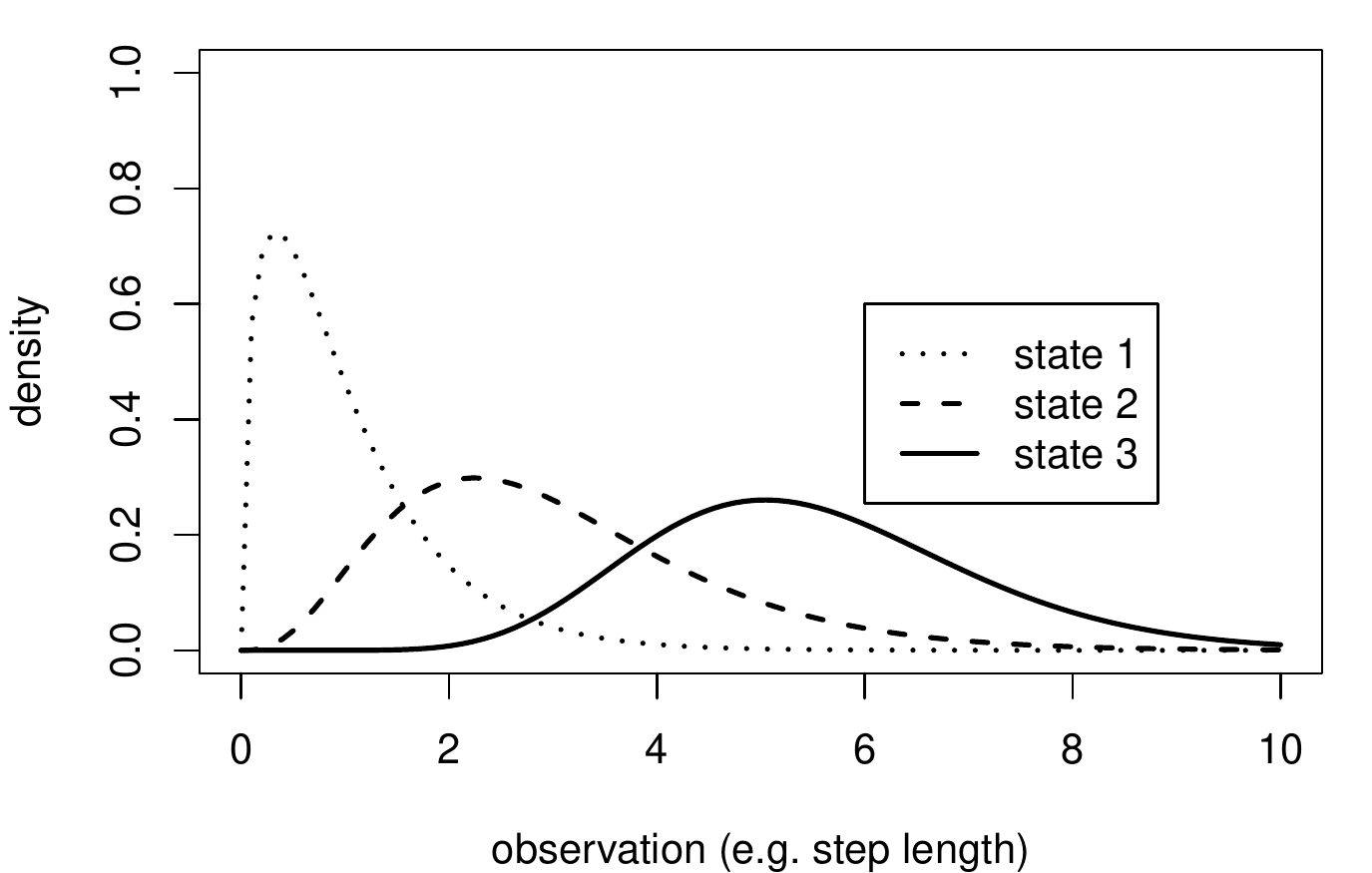}
\caption{Densities of the state-dependent distributions specified in Scenario 10.}\label{fig10}
\end{figure}

The state-dependent distributions specified in the two scenarios are displayed in Figures \ref{fig9} and \ref{fig10}, respectively. We note that especially Scenario 10 seems potentially biologically realistic --- for example, \citet{via17} find similar patterns for dive durations of harbor porpoises. For each of the two scenarios considered, Table \ref{tab:icsim2} displays the proportions of the 100 runs in which models with 2, 3 or 4 states were favored by AIC/BIC/ICL, respectively, when fitting the correctly specified models to the simulated data. In Scenarios 9 and 10, we find that the ICL does indeed tend to favor models with too few states, essentially confirming the findings by \citet{cel08} and in \citet{bac14}. While the AIC performed equally poorly as the ICL --- often selecting four and hence too many states --- the BIC almost always picked up the correct number of three states.

\begin{table}
\centering
\begin{tabular}{lcccc}		
	simul.\ 		& 	& \multicolumn{3}{c}{number of hidden states selected} 	\\ \cmidrule{3-5} 
 scenario			& criterion &	2 (\%)	&	3 (\%)	&	4 (\%)	\\\midrule
9                	& AIC		&	--		&	59		&	41		\\[0.2em]
					& BIC		&	2		&	98		&	--		\\[0.2em]
              		& ICL		&	64		&	36		&	--		\\ \midrule
10              	& AIC		&	--		&	54		&	46		\\[0.2em]
					& BIC		&	--		&	100		&	--		\\[0.2em]
              		& ICL		&	62		&	36		&	--		\\ 
%
\bottomrule
\end{tabular}
\caption{Percentages of runs in which the models with 2--4 states are chosen by AIC, BIC and ICL, for the two additional simulation scenarios 9 \& 10.}\label{tab:icsim2}
\end{table}

\renewcommand\refname{REFERENCES}
\makeatletter
\renewcommand\@biblabel[1]{}

\end{spacing}
\end{document}